# Inferring transport characteristics in a fractured rock aquifer by combining single-hole GPR reflection monitoring and tracer test data

Caroline Dorn[1], Niklas Linde[1], Tanguy Le Borgne[2], Olivier Bour[2], Maria Klepikova[2]

[1]Université de Lausanne, Centre de recherche sur l'environnement terrestre, Lausanne, Switzerland;

[2]Université Rennes 1-CNRS, OSUR Géosciences Rennes, Campus Beaulieu, 35042 Rennes cedex France.

**Inferring transport characteristics in a fractured rock aquifer by combining single-hole GPR reflection monitoring and tracer test data**

Caroline Dorn[1], Niklas Linde[1], Tanguy Le Borgne[2], Olivier Bour[2], Maria Klepikova[2]

[1]Université de Lausanne, Centre de recherche sur l'environnement terrestre, Lausanne, Switzerland;

[2]Université Rennes 1-CNRS, OSUR Géosciences Rennes, Campus Beaulieu, 35042 Rennes cedex France.

## Abstract

Investigations of solute transport in fractured rock aquifers often rely on tracer test data acquired at a limited number of observation points. Such data do not, by themselves, allow detailed assessments of the spreading of the injected tracer plume. To better understand the transport behavior in a granitic aquifer, we combine tracer test data with single-hole ground penetrating radar (GPR) reflection monitoring data. Five successful tracer tests were performed under various experimental conditions between two boreholes 6 m apart. For each experiment, saline tracer was injected into a previously identified packed-off transmissive fracture, while repeatedly acquiring single-hole GPR reflection profiles together with electrical conductivity logs in the pumping borehole. By analyzing depth-migrated GPR difference images together with tracer breakthrough curves and associated simplified flow- and transport modeling, we estimate (1) the number, the connectivity and the geometry of fractures that contribute to tracer transport, (2) the velocity and the mass of tracer that was carried along each flowpath, and (3) effective transport parameters of the identified flowpaths. We find a qualitative agreement when comparing the time evolution of GPR reflectivity strengths at strategic locations in the formation with those arising from simulated transport. The discrepancies are on the same order as those between observed and simulated breakthrough curves at the outflow locations. The rather subtle and repeatable GPR signals provide useful and complementary information to tracer test data acquired at the outflow locations and may help to characterize transport phenomena in fractured rock aquifers.

# 1. Introduction

Security concerns about waste disposals (nuclear, toxic waste, $CO_2$) and the need for efficient and sustainable extractions of natural resources (water, oil, gas, heat) in fractured rock formations require both process understanding and characterization of transport properties in fractured media. This implies the development of reliable monitoring technology for tracking temporal changes in the subsurface, in particular those related to contaminant transport. The limited accessibility to fractured rock systems contrasts with hydrological properties that are typically extremely heterogeneous at all scales [e.g. *Bonnet et al*., 2001; *Long et al.,* 1996; *Paillet,* 1998]. Generally, the data available for constraining fractured rock models have rather low information content with respect to the complexity of the system. For example, breakthrough curve data can be explained by a relatively small number of model parameters, while a very complex structure might have given rise to the observed data [e.g., *Becker and Shapiro*, 2000; 2003].

Models of conservative solute transport in fractured media typically combine advective and dispersive transport mechanisms within fractures with possibly matrix diffusion and sorption [*Maloszewski and Zuber*, 1985; *Hadermann and Heer,* 1996; *Lapcevic et al.,* 1999]. While these basic mechanisms are well known, a major challenge for modeling a system is the adequate description of heterogeneity at different scales. At the scale of the fracture, heterogeneous advection or "flow channeling" has been shown to be very common. Flow channeling, which is a phenomenon that increases in importance with the statistical variability in fracture aperture, refers to the situation in which the flow within discrete pathways make up a very large fraction of the total flow [*Tsang and Neretnieks, 1998*; *Moreno and Tsang*, 1994]. Highly localized fluxes and diffusion into stagnant inter-channel spaces within fracture planes are typically not taken into account in classical dispersion theories and the testing of alternative transport models requires detailed experimental investigations and imaging at the fracture scale [*Becker and Shapiro*, 2000].

Stochastic continuum methods can provide equivalent distributed models explaining observed state variables (e.g., temperature, pressure, tracer concentration, etc.) [*Neuman and Di Frederico*, 2003]. The applicability of such models at the local field-scale (1-100 m) is questionable in fractured rock systems as it is uncertain if a representative elementary volume (REV) exists as fractures often prevail at all scales [*Long et al.*, 1982; de *Dreuzy et al.;* 2001; 2002, *Neuman*, 2005]. Alternative representations based on Discrete Fracture Networks (DFN) [e.g., *Darcel et al*., 2003] are relatively difficult to condition and calibrate even with detailed measurements of aperture in boreholes or in-situ flow properties [*Neuman*, 2005]. Instead of building complex distributed models of heterogeneity, it is possible to account for unresolved heterogeneity and processes using effective models, for example, using concepts of multi-rate mass transfer [*Haggerty and Gorelick*, 1995; *Cvetkovic and Haggerty*, 2002], continuous time-random walk [*Berkowitz et al., 2006*], or multiple flow channels [*Becker and Shapiro*, 2000]. However, it is often difficult to assess which interpretive framework is the relevant considering typically available breakthrough data (or other hydrological data) [*Haggerty et al*., 2000, 2001; *Harvey and Gorelick*, 2000; *Le Borgne and Gouze*, 2008]. Obtaining spatially distributed images related to tracer movement within the formation can therefore be of key importance for defining appropriate effective models for transport in fractured media.

When analyzing breakthrough curves alone, it is generally not possible to uniquely (1) determine if transport occurs through one or several fractures and if multiple arrivals are caused by fracture heterogeneity (aperture variations) or by multiple flow paths involving different fractures or (2) infer what may be the cause of low mass recovery (e.g., through flowpaths driven by density effects or ambient flow; storage close to the injection point, in the fractures taking part in the tracer transport or through mass exchange with the rock matrix).

Geophysical imaging may provide information about subsurface structure and dynamics in-between the injection and extraction points, that is, at locations where hydrological data are generally not available [e.g., *Rubin and Hubbard*, 2005]. One of the most suitable geophysical methods in fractured rock investigations at the 1-100 m scale is ground penetrating radar (GPR). This method allows detecting mm-aperture fractures and resolving temporal changes away from the observation points [*Olsson et al.,* 1992; *Lane et al.,* 1996; *Lane et al.,* 1998; *Becker and Tsoflias*, 2010; *Dorn et al.*, 2011]. Surface GPR is useful to study transport in shallow subhorizontal fractures [*Talley et al.*, 2005; *Becker and Tsoflias*, 2010]. For larger depths, cross-hole difference-attenuation radar tomography [e.g., *Liu et al.,* 1998; *Day-Lewis et al.*, 2003] can image tracer movement through fracture zones, but the resolution of the resulting tomograms is insufficient for imaging transport in individual mm-aperture fractures.

*Dorn et al.* [2011] showed that time-lapse single-hole GPR data acquired during and after saline tracer injection tests allow imaging tracer movement through a network of connected fractures. The recovered images are relatively subtle despite extensive processing and many different time-lapses are necessary to make robust interpretations concerning transport pathways. Nevertheless, the resulting information about tracer transport and storage cannot be obtained by any other field technique that we are aware of, which warrants further study with this type of data. Herein, we build on the work by Dorn et al. [2011] by analyzing five tracer experiments (one of them being the experiment presented in our previous study) that were acquired under different injection and pumping conditions in a granitic rock aquifer. The objectives of this work are to show that time-lapse single-hole GPR reflection data acquired during saline tracer injection experiments make it possible to (1) obtain repeatable results, (2) identify transport pathways over tens of meters through connected individual fractures, (3) identify main transport mechanisms and causes of incomplete mass recovery at a site, and (4) provide geometrical constraints for the estimation of effective transport properties, namely hydraulic conductivities and dispersion coefficients. It is our hope that this contribution will motivate further research in how time-lapse GPR data can be used in fractured rock hydrology for (1) model validation, (2) model calibration, and (3) inversion processes. A simplified flow- and transport model calibrated to the breakthrough data is used to highlight some of these possibilities and associated challenges.

## 2. Methods

### 2.1 The single-hole ground penetrating radar reflection method

Ground penetrating radar (GPR) is an electromagnetic imaging method of the subsurface that is presented by *Annan* [2005], while *Balanis* [1989] describes the underlying physics of radar wave propagation. A GPR transmitter sends a source signal out into the medium, while a GPR receiver collects the resulting signals arising from signal transmission, reflections, and scattering at electromagnetic boundaries. The single-hole GPR configuration refers to the case in which the transmitter and receiver are both located in the same borehole at a known separation (see Figure 1). In single-hole reflection mode, imaged reflectors can arise from fractures located in all directions from the borehole as illustrated in Figure 1; planar reflectors that intersect the borehole are imaged as V-shaped reflections. Data processing of the acquired data allows determining the distances to the reflectors and their associated dips, but not their azimuth. Reflectors are predominantly related to variations in electrical permittivity $\varepsilon$, but also in electrical conductivity $\sigma$, and in magnetic permeability $\mu$. The attenuation of the signal propagating in the medium is proportional to $\sigma$. The data recorded by a receiver located at a given distance from the transmitter is traditionally used to image boundaries that in fractured media correspond to fracture surfaces [e.g., *Olsson et al.*, 1992; *Liu and Sato*, 2006]. When saline tracer arrives at a fracture, the locally elevated conductivity leads to increases in the reflectivity of the fracture and thus a higher amplitude GPR reflection. [*Tsoflias and Becker,* 2008]. An unwanted effect associated with measurements in boreholes following tracer injection tests is that temporal variations in fluid conductivity within the pumping borehole changes the radiation characteristics of the antenna [*Ernst et al.,* 2006] and therefore the effective source wavelet, which complicates the subsequent data processing.

Beside the medium constitutive parameters, the recorded reflection amplitude from a fracture depends on: (1) the fracture aperture and signal wavelength: closely spaced reflections from the upper and lower fracture surfaces interfere with each other [see *Tsoflias and Becker,* 2008; *Widess*, 1973]; (2) the dip of the fracture: the reflection coefficient of a dipping interface (or fracture) is a function of the signal angle of incidence and signal polarization [*Bradford and Deeds*, 2006; *Tsoflias and Hoch,* 2006], subvertical dipping features have higher reflection amplitudes than subhorizontal fractures when using GPR in vertical boreholes (0-30° dipping fractures are not directly detectable); (3) the distance between a fracture and the antennas: due to signal attenuation fractures are detectable up to roughly $r$ = ~15 m radial distance in granitic formations using a central signal frequency of 140 MHz; and (4) the spatial extent of a fracture: a reflection is an integration over an area of about the first Fresnel-zone (e.g., for a central frequency of 140 MHz, the Fresnel-radius is 0.6 m at a radial distance $r = 2$ m and 2 m at $r = 20$ m); (5) the azimuth of a fracture: reflections from a plane fracture can only be observed if a normal-vector to the reflector crosses the borehole; and (6) the roughness of the fracture: roughness creates diffractions that might allow imaging of fractures with unfavorable orientation.

## 2.2 Field site

The experiments presented herein were carried out within a fractured rock aquifer that constitutes the main water supply for the town of Ploemeur, France (Figure 2), with an average extraction rate of 2,000 L/min [*Le Borgne et al.,* 2006]. Our tracer tests were conducted ~3 km away from the water extraction site at the test-site Stang-er-Brune [*Le Borgne et al*., 2007]. The experiments were carried out between two ~6 m spaced boreholes B1 (83 m deep) and B2 (100 m deep). The boreholes reach a contact zone at a depth $z = $~40 m ($z = 0$ m corresponds to the top of the B1 borehole casing) between highly deformed micaschists and underlying saturated granite. Within the granite (at $z = 40$-80 m), the strongly deviated B2 is located 40-100°N relative to B1. The granite formation has the most permeable fractures [*Le Borgne et al.,* 2007] and is therefore the area of primary interest in this study (Figure 2).

*Le Borgne et al*. [2007] used televiewer data together with hydraulic testing (notably single-hole and cross-hole flowmeter tests) at the site to characterize fractures that intersect the boreholes and identify those that are hydraulically connected. The formation is highly transmissive with overall hydraulic conductivities on the order of $10^{-3}$ m$^2$/s over the length of each borehole. *Le Borgne et al*. [2007] reported an ambient vertical upward flow in the boreholes of about 1.5 L/min. This ambient vertical flow is the result of a 50 cm hydraulic head difference between the deepest fractures at $z = 100$ m and the upper micaschist. This regional upward flow that appears in all permeable boreholes is also expected to affect the well-connected fractures. The transmissive fracture network at the site is dominated by a relatively limited number of well-connected fractures (i.e., only 3-5 such fractures intersect a borehole over its entire length). These fractures have a dip in the range of 30-80° and an azimuth in the range of 190-270°. The dips and azimuths of the boreholes suggest that there is no single fracture that intersects both boreholes B1 and B2 [*Le Borgne et al.,* 2007].

*Dorn et al.* [2012] acquired 100 MHz and 250 MHz multifold single- and cross-hole GPR reflection data to constrain the geometry of the main fractures within the granite formation. Using the single-hole 250 MHz data, it was possible to obtain high-resolution images of the main fractures in the granite at radial distances $r = 2$-13 m away from B1 and B2 [*Dorn et al.,* 2012] including those that were identified as being transmissive by *Le Borgne et al.* [2007].

### 2.3 Experimental setup

Table 1 provides the experimental details of the tracer tests (referred to as experiments Ia, Ib, II, IIIa, and IIIb in the following) that were performed between B1 and B2 in June 2010. Figure 2c is a sketch of the experimental setup for the case in which B1 is the injection and B2 the pumping borehole. All logging takes place in the pumping borehole. Note that our naming convention is different than the one used by *Le Borgne et al.* [2007] in that we name each fracture according to the borehole name and the depth at which it intersects. For example, a fracture intersecting borehole B1 between 44.0 m $\leq z <$ 45.0 m is named B1-44.

For each experiment, a saline solution of ~90 L was injected during a short time interval (10-30 min) at a controlled rate into a transmissive fracture (experiments Ia and Ib in B1-78, experiment II in B1-50, and experiments IIIa and IIIb in B2-55) that was isolated from the rest of the injection borehole by a double-packer system. The initial tracer salinity was ~30 times higher than the background salinity of the groundwater. After the injection, we continued in four of the five experiments to push the tracer with fresh groundwater at approximately the same rate. For experiment Ib, no further injection of fresh groundwater was pursued after the end of the tracer injection. To pull the tracer solution towards the pumping borehole, we pumped water in the upper cased section of the pumping borehole. Salt concentrations were monitored below the pump at $z = 10$ m using an electrical conductivity logger. Although the mean transfer time between the two boreholes was about 1 to 3 hours depending on the experiments, pumping lasted for at least 12 hours to remove most of the tracer from the rock formation. Along the observation depth interval in the pumping borehole, we repeatedly acquired single-hole GPR data while measuring the borehole fluid electrical conductivity $\sigma_w$ and hydraulic pressure $p$ (one CTD logger was attached to the GPR antenna cable just above the upper antenna, Figure 2c). We used 250 MHz GPR antennas (MALÅ Borehole antennas with center frequencies around 140 MHz; antenna separation of 4 m) to obtain a high spatial resolution.

The different raw GPR sections $\mathbf{D}_i^{\mathrm{raw}}$ (depth sampling of $\Delta z = 0.1$ m) and corresponding $\sigma_w$ borehole logs (depth sampling of $\Delta z < 0.2$ m) were acquired over observation intervals of tens of meters. Each time-lapse $i$ is associated with an observation time $t_i^{\mathrm{obs}}$ relative to the start of the saline tracer injection. For each experiment, a reference GPR section $\mathbf{D}_1^{\mathrm{raw}}$ was acquired just before the injection, and the following sections $\mathbf{D}_i^{\mathrm{raw}}$ were acquired every 5-10 minutes (the acquisition of one GPR section takes approximately 5 minutes), except for the last section $\mathbf{D}_N^{\mathrm{raw}}$ that was acquired the following day after overnight pumping. Repeatability in the vertical positioning between the radar sections of a few cm were obtained by using a calibrated digital measuring wheel and by marking the start and end points on the cables. Two plastic centralizers attached to each GPR antenna assured that the lateral positions within the boreholes were similar between acquisitions (Figure 2c).

The whole suite of experiments (Table 1) allowed us to investigate under different conditions to what extent saline tracer transport in fractured media can be imaged with single-hole GPR reflection monitoring. The chosen injection points largely determine the fractures that take part in the tracer transport, but also variations in the injection and pumping rates will have a strong influence on the spreading of the tracer (especially at this site exhibiting significant ambient upward flow (~1.5 L/min) in the boreholes [*Le Borgne et al.*, 2007]). The main differences between experiments Ia and Ib were (as mentioned above) that no pushing of the injected tracer with groundwater was performed in experiment Ib and that the pumping rate was higher (~30 L/min) in experiment Ia than in experiment Ib (~16 L/min). The injection in experiment II was carried out in a fracture for which prior hydrological investigations indicate that the flowpaths toward B2 are rather subhorizontal. This is a challenging setup for single-hole GPR as subhorizontal fractures cannot directly be detected due to the high angle of incidence (tangential to the fracture) resulting in no reflected signal returning to the receiver antenna. Experiments IIIa and IIIb differ with respect to the previous surveys in terms of the higher injection rate (8-10 and 7-9 L/min); the pumping throughout experiment IIIa was unstable, whereas the pumping rate during experiment IIIb ranged between 5 and 6 L/min (see Table 1). *Dorn et al.* [2012] presented the GPR results from experiment IIIa and processing of experiment IIIb reveals similar results. The processing employed was slightly different than what is proposed below, but the overall tendencies were very similar. The results from the GPR processing presented herein therefore only consider experiments Ia, Ib, and II, while the results of experiments IIIa and IIIb are included in the interpretation.

**2.4 GPR data processing**

The most important aspect of successful GPR difference imaging is repeatability. Processing of high-frequency single-hole GPR reflection monitoring data is very challenging and a quite extensive testing of alternative processing strategies was necessary to assure that the difference amplitudes are comparable between acquisitions and to assure smooth transitions in the retrieved patterns between time-lapses. Indeed, positioning accuracy of sources and receivers is most important when imaging subhorizontally dipping fractures. Apart from standard GPR processing, we therefore had to account for (1) vertical positioning uncertainties on the cm-scale due to cable twisting during the data acquisition, (2) temporal variations in the effective GPR source signals caused by variations in the borehole fluid conductivity, and (3) significant direct wave energy and ringing signals caused by poor dielectric coupling that severely contaminate the individual raw sections $\mathbf{D}_1^{raw}$ to $\mathbf{D}_N^{raw}$ for traveltimes $t < 90$ ns (the direct wave is a wave traveling along the borehole wall). Generally, the raw data have high signal-to-noise-ratios for $t < 160$ ns.

Figure 3 summarizes the main processing steps of the GPR data. We accounted for time-zero drifts of the transmitter initialization time before correcting the residual misalignments of the direct wave between individual sections. An initial geometrical scaling of the signal was applied assuming spherical divergence of the source amplitude followed by a wide band pass filter in the frequency domain (linearly tapered with corner frequencies 0-20-300-380 MHz) that removes low and high frequency noise.

To minimize vertical positioning errors, we calculated depth corrections (Figure 3) using the processed data up to this point. To calculate the corrections, we aligned first-arrival energy, restricted the data to a time window after the first arrivals for times with a high signal-to-noise-ratio, applied a dip filter to suppress signals parallel to the direct wave, and narrowed the frequency spectrum of the data (60-70-190-210 MHz). We then calculated zero-crossing patterns of all data traces (1 for a zero-crossing before a maximum, -1 before a minimum and 0 otherwise). The vertical corrections were determined iteratively by searching, for each data trace, a correction that maximized the correlation between the zero-crossing patterns of an individual data trace and the corresponding stacked traces of all time-lapse data. The corrections were then used to construct a new stacked data section on which this process was repeated until the proposed correction from one iteration to the next was smaller than 3 cm on average. These corrections were applied to the widely bandpass filtered data (Figure 3).

To correct for temporal changes of the effective source signal due to salinity variations in the observation borehole, we followed *Dorn et al.* [2011] by applying a continuous wavelet transform and analyzing the wavelet power spectra of the data using the Morlet wavelet [*Torrence and Compo*, 1997]. In a first step, we removed wavelet scales with center frequencies outside the 20-160 MHz range. In a second step, we defined wavelet-scale dependent factors $\mathbf{F}_i$ as the ratios of the wavelet power of the direct wave of the processed data $\mathbf{D}_i^{\mathrm{proc}}$ with respect to the direct wave of the reference $\mathbf{D}_1^{\mathrm{proc}}$ ($=\mathbf{R}_1^{\mathrm{proc}}$). We then used the factors $\mathbf{F}_i$ to rescale $\mathbf{R}_1^{\mathrm{proc}}$ in the wavelet domain into new reference sections $\mathbf{R}_i^{\mathrm{proc}}$. The underlying assumption for this correction of the reference conditions is that the increased electrical conductivity of the borehole fluid affects the later arriving signals similarly as the direct wave, such that any remaining differences between time-lapses after this correction only reflect changes occurring within the rock formation. The reason for rescaling $\mathbf{R}_1^{\mathrm{proc}}$ instead of $\mathbf{D}_i^{\mathrm{proc}}$ is the higher bandwidth of $\mathbf{R}_1^{\mathrm{proc}}$ as high frequencies are strongly attenuated at later acquisition times due to the increasing borehole fluid conductivity.

To remove ringing signals caused by poor dielectric coupling, we applied an eigenvector filter that decomposes the data into eigenimages in a time window around the direct wave ($t < 90$ ns) using Karunen-Loeve theory. Then we excluded eigenimages representing ringing signals identified as those being parallel to the direct wave before reconstructing the data. After this preprocessing of the GPR data, the amplitudes are comparable and minimally affected by noise and signals other than reflections. As an example, the data at $t^{obs}$ = 45 min and its reference section of experiment Ib are shown in Figures 4a and 4b. It is important to note that the reflections corresponding to fractures (or changes in salinity within the fractures during the time-lapse experiments) are seen over a relatively wide time-window (e.g., the strong top reflector between $z$ = 40-50 m) and do not represent direct images of the fractures. In fact, the recorded GPR signal is a convolution of a finite source signal (~30 ns corresponding to ~3 m) with a rather discrete reflectivity distribution arising from the mm-aperture fractures. The time or distance to a given reflector corresponds to the first-arriving energy in these wave trains.

To facilitate the comparison of difference magnitudes, we calculated relative differences $\mathbf{M}_i$ (Figure 4c) over time by multiplying the differences $\mathbf{D}_i^{\text{proc}}$ - $\mathbf{R}_i^{\text{proc}}$ with the inverse envelope (reflection strength) sections of $\mathbf{R}_i^{\text{proc}}$. To avoid overinterpreting energy differences in low-reflectivity regions, we defined a minimum amplitude threshold for the envelope sections of $\mathbf{R}_i^{\text{proc}}$. Generally, the relative difference magnitudes vary smoothly between time-lapses. The largest changes occur during the first few time-lapses following the tracer injection and the signal generally returns towards zero at the end of the experiment (not shown). For experiments Ia and Ib it was necessary to further reduce the ambient noise level prior to migration (mapping data from time to distance from borehole) by applying a conservative f-x deconvolution (prediction filter in distance for each frequency) and subtracting a running median trace (Figure 4d, processing steps in brackets in Figure 3).

Pre-stack Kirchhoff depth-migration based on the 1-D velocity function of *Dorn et al.* [2012] made it possible to migrate $\mathbf{M}_i$ with minimal smearing or other artifacts (Figure 4e). Migration of difference sections is possible due to the linearity of migration with respect to the input wavefield term, which makes the final migrated sections comparable to migrated GPR sections [*Dorn et al.,* 2011]. The unmigrated difference sections $\mathbf{M}_i$ (Figure 4c) contain significant ambient noise at $t > 130$ ns, but the destructive superposition of ambient noise energy during migration significantly decreases the presence of incoherent events in the migrated images.

## 3. Results

### 3.1 Tracer test data

Figure 5 shows the measured electrical conductivities in the pumping borehole during the course of each experiment (interpolated from data logs with a depth sampling of $\Delta z < 0.2$ m and a time sampling $\Delta t \approx 10$ min, $\Delta t \approx 5$ min for experiment IIIb). The columns shown to the right of the each subfigure are the electrical conductivities acquired following overnight pumping. The flow and associated transport in the boreholes are directed upwards partly due to the natural upward gradient, but mainly because we pump at the top (except for experiment IIIa when the overnight pumping became weak and eventually stopped resulting in tracer accumulation at the bottom of the borehole).

The variations of electrical conductivity in time and space can be used to identify tracer and freshwater outflow zones by identifying those locations in which the electrical conductivity varies sharply in the vertical direction over extended time periods. These zones highlighted in Figure 5 correspond to open fracture locations in the optical logs; most of them identified by *Le Borgne et al.* [2007]: (1) B1-44, B1-50, B1-60 and B1-78; and (2) B2-49, B2-52, B2-55, B2-58 and B2-79. In experiments IIIa and IIIb, the location of the tracer outflow zone at the lower boundary of the observed depth interval (B1-78; see Figure 5d and e) is inferred from flowmeter data, as this is the only permeable zone below the observation interval. In order to identify the actual fractures through which tracer outflow occurs, we normalized the electrical conductivities (Figure 5) by the vertical flow distribution (Figure 6) of the fractures. In fact, some outflow zones do not carry significant amounts of tracer (see fractures B2-49 and B2-52 in Figure 5a-b, B2-58 and B2-79 in Figure 5c, and B1-44 and B1-52 in Figure 5d-e) and the tracer outflow zones B2-49 and B2-52 (experiment II) were not detected in the flowmeter data analysis of *Le Borgne et al.* [2007].

The peak electrical conductivities in the pumping borehole reach ~5 % of the injected tracer conductivity, except for experiment II where we observe only ~2.5 %. Such low percentages are due to (1) dispersion of the solute within the fractured media, (2) the pumping of fresh water that mixes with the saline water within the pumping borehole and (3) the influence of the ambient flow regime that may lead to tracer mass loss.

To estimate the curves of mass recovered at each individual fracture, we subtracted the estimated mass flux in the borehole below the fracture from the estimated mass flux above. When converting concentrations to mass rates, we accounted for the monitored, but rather unstable pumping rates (see Table 1), changes in pumping rates and available flowmeter data (Figure 6) that provide the relative contribution to flow of each outflow zone. To obtain the local mass recovery estimates, we then integrated the fracture-specific solute fluxes during the course of the GPR monitoring. The local mass recoveries should be analyzed with some caution as (1) the pumping rates and flow partitions between fractures are not perfectly known and (2) electrical conductivity logs acquired within the first 3-5 hours following the tracer injections do not capture the whole tail of the tracer breakthrough.

The derived fracture-specific curves of solute fluxes show very different characteristics for each type of tracer experiment (Figure 7).

(1) For experiments Ia and Ib (Figures 7a and 7b), the tracer injected in B1-78 arrives after 20 min in B2-79. Additional tracer arrival occurs after 30 min in B2-55 and B2-58 (subtle increase of salinity). Nearly half of the recovered mass comes from B2-79, while ~40

% of the recovered mass arrives at B2-55. After the main peak at $t^{obs} = $ ~45 min at B2-79, there is a second peak in the solute flux curve after ~90 min. At the end of the experiment, the total mass recovery is ~25 % for both experiments.

(2) For experiment II (Figure 7c), the tracer injection in B1-50 gives after 30 min rise to tracer breakthrough in B2 at B2-49 and B2-52. Additional tracer arrival occurs after 60 min in B2-55 and a very small amount of tracer arrives after 2.5 hours in B2-58. The fracture B2-49 contributes with ~20 % of the recovered mass, the largest contribution to the total mass (~50 %) comes from B2-52. None of these two fractures were detected by *Le Borgne et al.* [2007]. About ~20 % of the recovered mass arrives at B2-55. The estimated total mass recovery of salt after overnight pumping is ~30 %.

(3) For experiments IIIa and IIIb (Figure 7d and 7e), the tracer injection in B2-55 gives after only 10 min rise to tracer breakthrough in B1-60. Within the first 2 hours, ~40 % of the recovered tracer mass arrives at this fracture. After one hour, tracer arrival occurs in B1-78, where ~60 % of the total mass is recovered. The total mass recovery at the end of the GPR acquisition experiment is ~20% for experiment IIIa and after overnight pumping ~30% for experiment IIIb.

In summary, (1) the estimated solute fluxes from the outflow zones in the pumping borehole are the result of several different fractures or pathways; (2) the mass recoveries are generally relatively low, which we attribute to the ambient flow regime, density effects and the injection conditions.

## 3.2. Single-hole GPR data and difference imaging

Figures 8 and 9 show, for each of the experiments Ia, Ib and II, five migrated GPR relative-difference sections acquired at different times plotted on top of the migrated GPR sections of *Dorn et al.* [2012]. These images represent changes in reflection amplitudes in a 2-D projection around the borehole (i.e., depth $z$ and radial distance $r$) with respect to background conditions. Note that the GPR difference-images have a large imprint of the finite source signal. This implies that the fractures in which salinity changes occur are located where the first arriving energy of the reflection wave trains are observed. Figure 10 highlights the reflections in the background images that correlate to the high magnitude patterns in the difference sections (Figures 8 and 9).

Patterns of high magnitudes have predominantly subhorizontal to vertical dips ranging from 30° to 90° (relative to the surface) covering radial distances $r$ = 2-10 m from the pumping borehole. It is not possible to resolve features for $r < 2$ m due to the very high direct wave amplitudes in the corresponding time interval that completely mask the much smaller reflection amplitudes. Subhorizontal patterns show overall weaker amplitudes than subvertical patterns, which can be attributed to the high angle of incidence to the fracture surface. Difference patterns close to the injection point are predominantly imaged with high magnitudes at early times $t^{\text{obs}}$ (Figures 8a, 8f, 9a). At depths different than the injection point, patterns with typically weaker amplitudes appear at later $t^{\text{obs}}$ and are generally visible for longer time periods. Patterns of evolving difference magnitudes can generally be traced from the injection points through depth intervals ranging over some tens of meters. The GPR difference images in Figures 8 and 9 are discussed in detail below together with some of the trends observed at the many intermediate observation times that are not shown (see Table 1).

### 3.2.1 Experiments Ia and Ib

The main difference in experimental design between experiments Ia and Ib is that we did not push the injected tracer with groundwater in experiment Ib after the end of the tracer injection. During tracer injection, pumping conditions were similar for both experiments. We expect significant differences for the tail of the solute flux curves if we push the tracer with groundwater or not, but less for the mean arrival times that are similar for the two experiments (Figures 7a-b). The spatial distributions of the GPR magnitude patterns and their evolution over time are similar for the two experiments (Figure 8) (the difference amplitudes are plotted with the same scale in all sub-figures). At early times (Ia: $t^{obs}$ = 10 min, Ib: $t^{obs}$ = 15 min; Figures 8a and 8f), a single high magnitude pattern is visible at $z$ = 75-80 m indicating an upward movement of the tracer from the injection point towards the pumping borehole. At later $t^{obs}$ (Figures 8b, 8c, 8g and 8h), new magnitude patterns with dips between 40° and 80° show up in this depth region. The magnitudes get weaker with time and patterns start to appear at shallower depths (up to ~50 m depth). Two hours after injection, magnitude patterns for experiment Ia appear patchy and weak (Figure 8d), while for experiment Ib they are continuous and moderate in amplitude (Figure 8i). The next day acquisition for experiment Ia (Figure 8e) does not show weak and random amplitudes as experiment Ib (Figure 8j), but a region of moderate amplitudes above the injection point. These remaining amplitudes are partly attributed to unstable overnight pumping.

By overlaying the difference images on the migrated GPR sections of *Dorn et al.* [2012], we find at least 9 fractures through which the tracer solution likely moved (highlighted fractures in Figure 10a). Close to the injection point there are three prominent fractures through which the tracer moves upward (dipping 70°, 75° and 80° between $r$ = 3.5 and 7 m). The imaged magnitude pattern close to the pumping borehole at $r$ = 2-4 m and $z$ = 76 m, which is dipping 40° is most likely related to the fracture through which the tracer flows into the pumping borehole in B2-79 (Figures 7a-b, 8a). Between $z$ = 50 m and 70 m, the tracer solution moves through at least 4 more fractures, but the connections between these fractures are not always clearly imaged. Most probably, fractures outside of the detection range (with respect to dip and azimuth) carry some of the tracer. Even though tracer arrival at $z$ = 79.3 m has been imaged, it is clear from the difference images that the tracer mainly moves upward through a network of connected fractures. This upward movement corresponds well with the later arriving tracer in B2-55 and B2-58 as shown in Figures 5a-b. The fractures carrying the tracer into the borehole in this depth interval are probably not imaged because of the low dips. In fact, optical logs indicate dips of 33° and 31° for these fractures.

We attribute the less patchy and stronger relative-difference magnitudes in experiment Ib to the tracer solution being spatially more contained compared with experiment Ia. The pushing of the tracer in experiment Ia might have partially pushed the tracer in other directions than the pumping borehole B2, and therefore increased the spreading of the tracer. The overall similarity of the two sets of difference images representing experiment Ia (Figures 8a-e) and experiment Ib (Figures 8f-j) makes us confident (considering experimental differences) that the experiments are generally repeatable and the processing scheme can handle the majority of the experimental uncertainties. Furthermore, the dips and locations of the patterns correlate well with previously imaged fractures using multi-offset single-hole data [*Dorn et al.,* 2012] and hydrogeological studies [*Le Borgne et al*., 2007]. It appears thus that we can identify the main tracer-occupied fractures by superimposing migrated relative difference sections on the migrated multi-offset single-hole GPR data. Nevertheless, the interpretation must consider many intermediate acquisition times to assure that interpreted features are not related to processing artifacts. The interpretation shown in Figure 10a is based on careful analysis of the 16-21 difference-images obtained for each experiment (see Table 1).

### 3.2.3. Experiment II

The migrated difference images from experiment II (Figure 9) show complex magnitude patterns of superimposed reflections that are limited to the depth range of $z$ = 45-60 m. The superposition of different signal contributions and the shallow dips of the fractures through which the tracer appears to move make it very difficult to trace magnitude patterns related to individual fractures. One well-resolved feature is the spatially compact high amplitude pattern above and behind the injection point that is visible at early times $t^{obs} < 50$ min ($z = 47$-$50$ m and $r = 5.5$-$7.5$ m in Figure 9a-b). Polarity changes are found at successive observation times at $z = 49$ m (Figure 9a-b), which appear to cut horizontally through this subvertical reflectivity pattern. Rather complex weak-to-high magnitude patterns at smaller radial distances are also seen in Figure 9a-c. At $t^{obs} > 50$ min, weak magnitudes appear at $r = 9$-$10$ m in the depth range $z = 45$-$48$ m. After 2 h, the magnitude patterns between the injection and pumping boreholes in the depth range $z = 45$-$52$ m have weakened considerably and the remaining magnitudes are mostly concentrated in a region below the injection point (Figure 9d). The data acquired on the following day does not show any significant remaining relative-difference magnitude patterns, thereby indicating that most of the tracer has left the observable region.

The small differences between the tracer injection depth in B1 ($z = 50.9$ m) and the outflow depths in B2 ($z = 49$-$58.9$ m) makes the GPR interpretation difficult as it suggests that flow paths are rather subhorizontal. We find that the difference patterns are largely limited to a region in-between the injection and pumping borehole covering a similar depth interval as the outflow locations. The pattern that is imaged close to the pumping borehole (dipping 30°) around $z = 48$ m is likely related to B2-49 at which the tracer arrives in the borehole. The polarity changes discussed above are also observed in the migrated GPR sections of *Dorn et al*. [2012]. They might originate from nearly horizontal reflection boundaries in-between the boreholes. We have been conservative in interpreting these polarity changes in the difference images as possible tracer transport paths and only indicated one where we could correlate it to the GPR sections by *Dorn et al.* [2012] (see Figure 10b). A possible explanation for the low mass recovery of this experiment is that this horizontal fracture at $z = 48$ m that can be traced over $r = 4\text{-}10$ m carried the tracer away from the injection point to a larger subvertical fracture zone located at larger radial distance ($r = 9\text{-}10$ m) at $z = 45\text{-}50$ m.

### 3.2.4. Experiments IIIa and IIIb

Experiments IIIa [*Dorn et al.*, 2011] and IIIb (not shown due to strong similarity with the results of experiment IIIa) indicate a strong influence of the natural gradient. By interpreting the overlaid images, we observe at least 6 tracer-occupied fractures (Figure 10c), including a large fracture zone covering a wide depth interval of $z = 40\text{-}65$ m [c.f., *Dorn et al.*, 2011]. The saline tracer quickly moves down through two fractures dipping 50° and 75°. The tracer arrives in the pumping borehole through fracture B1-60. Tracer arrivals at greater depths cannot be inferred by our difference images alone as the tracer outflow occurs close to the bottom of the pumping borehole. It is likely that fracture B1-78 carries the tracer to the borehole. This fracture appears to be directly connected to the fractures we observe at $z > 70$ m in experiments Ia-b (Figure 10a).

The results in this section clearly demonstrate that the GPR difference patterns are related to transport within connected fractures. Evidence for this is given by the gradual spreading of the GPR difference-patterns away from the injection point (Figures 8 and 9; Figure 1 in *Dorn et al.* [2011]), the similarity of the inferred patterns for repeat experiments that include the same injection fracture (experiments Ia and Ib in Figure 8; experiments IIIa and IIIb (not shown)), and an overall agreement between the timing of the depth intervals experiencing temporal changes in the GPR images (Figures 8 and 9; Figure 1 in *Dorn et al.* [2011]) and the arrival of saline tracer at the outflow locations (Figure 7).

## 4. Comparison of GPR reflection sections with tracer transport modeling

In this section, we investigate in a more quantitative manner the agreement between tracer transport and the GPR difference sections. To do so, we calibrate a fracture model representing experiment Ib using a simplified three-fracture model that only models the main features of the observed mass flux curve in Figure 7b. This 3D model (with 2D flow in each fracture) is simplistic in that it (1) ignores the azimuths of the fractures, (2) it merges several connected fractures in one large fracture, (3) no heterogeneities of the fractures are considered except for classical dispersion parameters, (4) no density effects are considered, and (5) the natural gradient is ignored. This presented model is clearly very simplified, but it is useful to assess if changes in the difference sections at chosen locations are consistent with the simulated tracer distributions. We investigate below if discrepancies between the inferred curves are similar to those observed for the simulated and observed solute transport at the outflow locations. If this is the case, we argue that the resulting GPR-inferred reflectivity changes can be used to derive semi-quantitative breakthrough curves at locations between the boreholes.

The map of interpreted tracer pathways (Figure 11a) for experiment Ib is used as a basis to define continuous transport pathways between injection and outflow locations (Figure 11b). The model in Figure 11c combines the three transport pathways in Figure 11b into three fractures. The distances between outflow and injection locations are inferred from Figure 11b. The sketched 3D fracture planes in Figure 11c are modeled with an aperture of 1 mm and an extension of ±100 m in the out-of-plane dimension. We modeled flow and transport using COMSOL Multiphysics 3.5 using a finite element mesh with 6,200 elements. We solve Darcy's law with the observed time-varying head boundaries in the pumping well for the outflow locations (Figure 11e) and use mixed boundary conditions at the injection location (the observed fixed head during the injection period and zero flow conditions afterwards). The edges of the fractures are modeled as zero head boundaries. Using the calculated velocity field, we solve the advection-dispersion equation assuming a constant concentration at the injection location during the injection period. The free fitting parameters are hydraulic conductivities, dispersivity and the concentration as a fraction of the actual injected concentration. The latter allows us to partly consider mass loss and to fit the magnitudes of the observed solute fluxes.

The effective parameters for the three fractures were obtained by manual calibration aiming at fitting the first arrival times and the peak solute fluxes at $t^{obs} \sim 1.4$ hours. An automatic calibration procedure based on a Levenberg-Marquardt algorithm was also used, but did not provide significantly better results. It is clear that the simulated curves in Figure 11d (dotted lines) only represent some of the main characteristics of the measured curves (solid lines), which makes the estimated transport properties rather approximate. The derived effective hydraulic conductivities are $k_1 = 0.6$ m/s, $k_2 = 2.3$ m/s and $k_3 = 0.2$ m/s and the dispersivities are $\alpha_1 = 0.6$ m, $\alpha_2 = 0.3$ m and $\alpha_3 = 0.2$ m. The fitted curves underestimate mass fluxes at early arrival times and overestimate them after peak arrivals. Fluctuations in the pressure conditions significantly influence the shape of the modeled flux curves, for example, by reproducing observed peaks at $t^{obs} \sim 1.4$ and 2.3 hours (Figure 11e).

The resulting concentration fields were used to calculate how the simulated tracer distributions affect GPR reflectivity. This analysis is based on local tracer concentrations that correspond to the three locations highlighted in Figure 11b (green highlighted letters). The concentrations are mapped into $\sigma_w$ values that we relate to variations of reflection coefficients of thin-layer reflectors using the expression of *Deparis et al.* [2003] under the assumption of a normal incidence wave.

Reflectivity strengths from the GPR difference sections are retrieved at each depth location by picking the maximum value around the chosen location (marked by letters K-M in Figure 11b). Figure 12 plots the time evolution of these picked reflectivity strengths

(asterisks, normalized to the maximum of its fitted second order polynomial) and the estimated reflection coefficients from the simulated concentrations (solid line, normalized to the maximum value). Note that both estimates have been averaged over a 1 m large zone at each location.

Figure 12 illustrates that the times at which reflection strengths rise are overall similar for the two estimates. The earliest rise of the reflection strengths are observed for region K (6 m away from the injection point), which is consistent with the observed breakthrough data in that the first arriving mass is found along this flowpath. Reflection strengths are found to rise earlier in region L than in region M, which is consistent with their distance to the injection location (L is 8 m and M is 19 m away from the injection location). The main discrepancy between the curves is that the picked reflection strengths from the GPR difference-images start to go down after ~1-2 hours, which is not seen in the simulated reflectivity coefficients based on transport modeling, except for position K (Figure 12a). This is consistent with the discrepancy evoked earlier between the simulated and observed solute fluxes in Figure 11d. Indeed, the simulated tracer concentration stay relatively high at the end of the simulations, while both the GPR data and the estimated solute fluxes at the outflow locations indicate that the tracer concentration goes down significantly after the peak arrival.

The results in Figure 12 provide evidence that the amplitude changes in the GPR data are directly related to concentration changes within the fractures, which implies that we can obtain relative breakthrough curves for locations between the observation boreholes. Absolute breakthrough curves would require more precise knowledge of the GPR source signal or alternatively a calibration of the radar reflected amplitudes to controlled tracer concentrations as done by *Becker and Tsoflias* [2010]. Another striking aspect of Figure 12 is that the picked reflectivity strengths vary relatively smoothly over time. The fact that there is no smoothing applied to the GPR-inferred reflectivity changes over time gives confidence that the GPR data provide information about solute transport at locations within the formation.

## 5. Discussion

The GPR difference amplitude images presented in this work (Figures 8-9) and by *Dorn et al.* [2012] provide useful complementary information to classical breakthrough data. Although each method has its limitations, we argue that their combination have a high potential to improve characterization and lead to new insights about tracer transport in fractured media. The main limitations are: (1) breakthrough data provide integrated responses between injection and observation points; (2) the GPR data do not provide information about the region in the intermediate vicinity of the pumping borehole; (3) the GPR data are only 2D projections imaging those parts of the local fracture planes that have a favorable orientation with respect to the acquisition geometry, and they do not provide information on the azimuth of fractures.

For all tracer injection experiments, we find that multiple transport paths carry the tracer between the injection point and the pumping borehole. This is seen already by considering the distribution of mass rates along the borehole (Figure 7), but the GPR difference images offer a more complete view of fracture connections and transport pathways between the two boreholes (Figures 8-9; Figure 1 in *Dorn et al.* [2011]). In all experiments, we find that the depth intervals and the timing of the GPR magnitude patterns agree well with the calculated mass rates. We see similarly located magnitude patterns in experiments Ia and Ib (Figure 8), as well as in experiments IIIa and IIIb (not shown). The patterns appear at slightly different times due to differences in the experimental setup (e.g., in terms of the pumping and injection rates). The similarities of the two sets of difference images obtained from repeat tests using the same fracture for injection make us confident that the experiments are repeatable and the processing scheme is robust. Our results suggest that it is possible to identify the main tracer-occupied fractures over time by superimposing migrated relative difference sections on the migrated GPR sections from B1 and B2 acquired under natural flow conditions [*Dorn et al.,* 2012]. When comparing these images, it seems that the most prominent fractures imaged by *Dorn et al.* [2012] also carry tracer in the saline tracer experiments. The GPR difference images provide us with a plausible explanation about where the unaccounted mass went. For the experiments presented herein, it seems that most of the missing mass was transported from the injection point in a direction away from the pumping boreholes and the observable region, while some storage or delay in less mobile zones may have occurred.

When interpreting these results, it is important to consider that the GPR relative-difference images provide an incomplete description of tracer movement as the fracture azimuth is unresolved, but also because certain fractures that carry tracer will not be imaged. These include small-scale fractures (i.e., with a fracture surface smaller than about the first Fresnel zone (0.6 m at $r$ = 2m and 2m at $r$ = 20 m)), fractures with subhorizontal dips, fractures with an unsuitable azimuth, and fractures located close to the boreholes. We observe dips of fractures at all detectable angles (considering that only dips between 30°-90° are detectable), the most common dip being around 30°. The recorded GPR difference-amplitudes are surface-averaged measures (over the first Fresnel zone) of electromagnetic contrasts, which imply that the difference images have a limited sensitivity to tracer dispersion within a single fracture. Still, in fractures that are imaged in the difference sections it is very likely that spreading within the fractures occur at least on the meter scale.

Differences in the experimental setup lead to observable differences in the temporal and spatial dynamics of the tracer transport. First of all, the pathways but also the ratios between the imposed heads and the upward natural gradient differ: In experiments Ia and Ib with an injection rate on the order of the natural gradient, the tracer moves upwards and spreads over tens of meters (Figures 8 and 10a), in experiment II with again an injection rate on the order of the natural gradient, the tracer moves subhorizontally (Figures 9 and 10b); and in experiments IIIa and IIIb where the injection head is roughly three times stronger than the natural gradient, the tracer moves partly downwards and spreads over tens of meters (Figure 1 in *Dorn et al.* [2011] and Figure 10c). We observe multiple peaks in the solute flux inferred at different fracture locations (e.g., in B2-79 at $z = 79.3$ m for experiments Ia and Ib, Figure 5a-b) that are attributed to variations in the pumping rates as we see similar behavior in the simulated breakthrough curves of experiment Ib (Figure 11d).

For our experimental setup we have to note that the relative contributions to flow and mass of a given fracture depend on overall connectivity with the permeable fracture network, whereas the mass contribution depends on the local connections with the injection fracture. In experiment II, most of the mass arrives through a fracture that does not contribute significantly to flow (in B2 at $z = 52.7$ m). In fact, this fracture was not even identified by *Le Borgne et al.* [2007] when analyzing flowmeter data from the site. In experiments IIIa and IIIb, the recovered mass is arriving nearly in equal parts at two fractures located 22 m apart, one contributing with 80% and the other <5% to flow.

The mass recovery is low (<30%) in all experiments. The tracer might move out into fractures that carry the tracer away from the pumping borehole either due to the ambient flow field, by density effects or by the injection pressure. The pushing of the tracer by continued water injection in experiments Ia, II, IIIa and IIIb likely pushed some of the tracer away from the pumping borehole. In these cases, the pumping might only weakly affect the tracer and its subsequent movement. The regional upward gradient that is manifested by a ~1.5 L/min flow in the boreholes [*Le Borgne et al.*, 2007] seems to influence the tracer movement for some of the experiments, for which we observe significant upward movement of the tracer into larger fracture zones (experiments Ia-b and IIIa-b).

Tracer transport between the two ~6 m distant boreholes is fast for the experiments presented here. Tracer breakthrough occurs during the first hour in all experiments and peak concentrations in the borehole fluid are observed after 30 min (experiments Ia and Ib, Figure 5a-b) to 3 hours (experiment II, Figure 5c). Correspondingly, the GPR difference images evolve quickly in time during early observation times. The corresponding apparent tracer velocities considering the length of the 2D projected pathway between injection and outflow locations in Figure 10 and the minimal tracer travel time give estimates in the range of 0.2-1.3 m/min. These velocities of the first arriving tracer are likely higher as the actual travel path length in 3-D is larger.

A comparison between picked GPR reflectivity changes at specific locations over time with those inferred from flow and transport modeling (Figure 12) show a good agreement at early times and discrepancies at late times, indicating that a more complex flow and transport model could be constrained with these data. The discrepancies are on the same order as those between observed and simulated solute fluxes at the outflow locations using the same flow and transport model. The time-series of GPR reflectivity changes have a high signal-to-noise ratio and indicate not only the arrival time of the saline tracer at a specific location, but also how the tracer concentration decreases over time. Forced tracer tests examine only the fractures that are involved in tracer transport and does not represent natural conditions. To better understand and build models for predicting flow and transport under natural conditions one must carry out experiments under natural flow conditions. Using single-hole GPR difference imaging as presented here offer the possibility to image transport under such conditions even in the case of no or very limited tracer arrival in the boreholes.

The resulting GPR difference sections are a result of a rather extensive processing workflow. Research is warranted to better understand under what conditions this type of data can provide reliable information about transport within specific fractures and how to best use such data to constrain realistic 3-D fracture network models that honor not only borehole information, but also transport pathways, effects of natural flow gradients, and storage changes imaged by the GPR data. To facilitate the interpretation of the difference-migrated images, it would be most fruitful to test and further develop suitable deconvolution algorithms that remove the imprint of the GPR source signal [*Schmelzbach et al.,* 2011].

## 5. Conclusions

We find that time-lapse single-hole GPR data acquired during saline tracer injection tests provide insights about the temporal evolution of tracer plume geometry that is complementary to information derived from classical hydrological characterization of fractured aquifers. The GPR data make it possible to derive a length scale of the fractures involved in the tracer transport and to infer the connectivity and geometry of these fractures. Furthermore, the data help to better understand where the tracer that did not arrive in the pumping boreholes went. For five tracer experiments in a fractured granite, we find that the GPR data acquired with 250 MHz antennas provide subtle but reliable images of the evolution of tracer plumes through time at radial distances $r$ = 2-10 m from the boreholes (Figures 8-9). Hydrological data and migrated relative difference amplitude images derived from the GPR experiments are consistent with each other and indicate similar tracer transport characteristics for the experiments that involved the same injection fracture. For all experiments, we find that multiple pathways involving several fractures connect the injection fracture with the pumping borehole and that the total vertical spread of the tracer is in the range of tens of meters despite that the two boreholes are only located six meters apart. The vertical ambient pressure gradient at the site seems to carry most of the injected tracer upwards through fractures that do not intersect the pumping boreholes, while some storage of tracer mass appears to occur in less mobile zones within the inter-borehole region. We find that 2D geometrical information about pathway lengths and connections help to constrain breakthrough analyses. We demonstrate also for one of the experiments using a simplified fracture model how GPR reflectivity time-series at chosen locations may be used to test and provide further constraints to transport models.

### Acknowledgements

We thank the field crew of Ludovic Baron, Nicolas Lavenant, and Vincent Boschero. We are thankful to Alan Green at ETH-Zurich for making his GPR equipment available. This research was supported by the Swiss National Science Foundation under grant 200021-124571, the French National Observatory H+, the European ITN network IMVUL (Grant Agreement No. 212298), and the European Interreg IV project CLIMAWAT. Many useful comments from three anonymous reviewers and the AE Frederick Day-Lewis helped to improve the manuscript.

**Table captions**

**Table 1.** Experimental setup of the five tracer experiments. The listed pumping rates refer to the time periods of GPR monitoring.

**Figure captions**

**Figure 1.** (a) Principle of the single-hole GPR reflection method, in which a transmitter T sends out a signal that is reflected and subsequently collected by a receiver R located in the same borehole as the transmitter. (b) Schematic reflection section illustrating typical reflection patterns arising from intersecting and non-intersecting fractures. Figure from *Spillmann et al.* [2007].

**Figure 2.** (a) Location of the Stang-er-Brune study-site in the vicinity of Ploemeur, France. (b) Geological model of the field site with a 30° dipping contact between micaschist and underlying granite. (c) Schematic of the data acquisition setup, in which $p$ and $\sigma_w$ logger refer to hydraulic pressure and groundwater conductivity loggers, respectively.

**Figure 3.** Flowchart of the GPR processing steps with reference to figures showing intermediate results. Processing steps in parenthesis only apply to experiments Ia and Ib.

**Figure 4.** Results of data processing applied to a single-hole GPR data section of experiment Ib acquired after saline tracer injection at $z = 78.7$ m ($t^{obs} = 45$ min, 4 m antenna offset). (a) Data and (b) reference section after preprocessing. (c) Relative difference between (a) and (b) normalized by the envelopes of (b). (d) As in (c), but after f-x deconvolution to remove noise. (e) As in (d) but after depth migration using the velocity model shown to the right of the migrated difference. The axis aspect ratio $r$:$z$ is 2:1.

**Figure 5.** Electrical conductivity $\sigma_w$ in the pumping borehole during experiments (a) Ia and (b) Ib, (c) II, and (d) IIIa and (e) IIIb. The values of $\sigma_w$ the day after saline injections are shown in separate columns to the right of the subfigures. Black triangles mark the acquisition times of the conductivity profiles shown in Figures 8 and 9. The red arrowheads indicate locations with interpreted tracer outflow, while the blue arrowheads indicate outflow locations that are unaffected by the saline injections.

**Figure 6.** Induced vertical flow due to pumping in (a) B1 with 82 L/min and in (b) B2 with 138 L/min measured with an impeller flowmeter. All the flow below $z = 60$ m in B2 stems from fracture B2-79 at $z = 79.3$ m. Arrows indicate the locations of interpreted permeable fractures.

**Figure 7.** Local salt solute flux curves estimated at depth locations with significant tracer arrival (colored lines) and below the pump ($z = 10$ m, black line) normalized by the injected amount of tracer mass for experiments (a) Ia, (b) Ib, (c) II, (d) IIIa and (e) IIIb.

**Figure 8.** Migrated relative difference GPR sections acquired in B2 during experiment Ia (a-e) and experiment Ib (f-j) at different observation times $t^{obs}$, superimposed on the gray opaque migrated GPR section acquired under natural flow conditions [*Dorn et al.,* 2012]. High difference patterns originate from increased salinity in fractures located at the front of each such pattern (i.e., the smallest radial distance $r$ for each depth $z$). Note that we do not image any features at $r < 1.5$ m (gray region) because of the dominance of the direct wave at early times and its subsequent removal, which tends to remove superimposed reflections at early times. The corresponding electrical conductivities $\sigma_w$ of the borehole fluid in B2 are shown in the color profile at $r = 0$ m.

**Figure 9.** Migrated relative difference GPR sections acquired in B2 during experiment II at different observation times $t^{obs}$, superimposed on the gray opaque migrated GPR section acquired under natural flow conditions [*Dorn et al.,* 2012]. The corresponding electrical conductivities $\sigma_w$ of the borehole fluid in B2 are shown in the color profile at $r = 0$ m.

**Figure 10.** Extracts of the migrated multi-offset single-hole GPR sections of B1 and B2 from *Dorn et al.* [2012] with superimposed interpretations of tracer pathways for experiments

(a) Ia and Ib, (b) II, (c) IIIa and IIIb. Red circles indicate the tracer injection points, while red and blue arrowheads locate saline and unaffected groundwater outflow into the pumping borehole, respectively. Light red regions highlight fractures through which the injected tracer is interpreted to move, whereas blue regions highlight reflections from other boreholes. Light blue letters refer to transmissive fractures identified in the boreholes using optical logs and flowmeter tests with corresponding blue lines indicating their corresponding dips [*Le Borgne et al.,* 2007]. (f) Dip angles corresponding to the axis aspect ratio *r*:*z* of 2:1.

**Figure 11.** Transport model and modeling results for experiment Ib based on a simplified 3D fracture model with three intersecting rectangular fracture planes with an aperture of 1 mm. (a) Extract of Figure 10a, on which we assign (b) three continuous transport pathways between injection and outflow locations. Letters K-M in (b) refer to the positions considered in Figure 12. (c) Graph representing a simplified representation of the pathways in (b) used in the transport model to estimate effective hydraulic conductivities $k$ and dispersivities $\alpha$. (d) Observed (solid lines) and simulated (dashed lines) local salt flux curves at outflow locations B2-55, B2-58 and B2-79. The derived transport parameters are: $k_1 = 0.6$ m/s, $k_2 = 2.3$ m/s, $k_3 = 0.2$ m/s, $\alpha_1 = 0.6$ m, $\alpha_2 = 0.3$ m and $\alpha_3 = 0.2$ m. (e) Hydraulic head used for the boundary conditions at the three outflow locations.

**Figure 12.** Time evolution of normalized reflectivity changes with respect to reference conditions observed in the GPR difference images (asterisks) and calculated from simulated tracer distributions (solid line) corresponding to locations (a) K, (b) L and (c) M in Figure 11b. At each location, we average the data over 1 m, but there is no smoothing over time.

| Experimental parameters | Experiments | | | | |
|---|---|---|---|---|---|
| | Ia | Ib | II | IIIa | IIIb |
| | *Injection well* | | | | |
| Fracture of injection | B1-78 | B1-78 | B1-50 | B2-55 | B2-55 |
| Depth of injection | 78.7 m | 78.7 m | 50.9 m | 55.6 m | 55.6 m |
| Injection rate | 2-3 L/min | 2.3-3.5 L/min | 2.3-2.7 L/min | 8-10 L/min | 7-9 L/min |
| Amount of tracer | 87 L | 90 L | 92 L | 93.5 L | 92.5 L |
| Injected amount of salt | 3.5 kg | 4.7 kg | 3.7 kg | 4.7 kg | 4.6 kg |
| Injected amount of uranine | 3.5 g | 6.2 g | 4.1 g | 50 g | 10 g |
| Tracer conductivity | 5 S/m | 5.5 S/m | 5 S/m | 5.5 S/m | 5.5 S/m |
| Pushing tracer with fresh water | yes | no | yes | yes | yes |
| | *Observation well* | | | | |
| Borehole | B2 | B2 | B2 | B1 | B1 |
| Number of time steps $N$ | 16 | 21 | 29 | 31 | 33 |
| Observation interval | 35-95 m | 35-90 m | 35-85 m | 35-80 m | 35-75 m |
| Range of pumping rates | ~30 L/min | 5-30 L/min | 13-25 L/min | 1-10 L/min | 5-6 L/min |
| Mean pumping rate | 30 L/min | 16 L/min | 16 L/min | 6 L/min | 5.5 L/min |
| Mean time step | 10 min | 10 min | 10 min | 10 min | 5 min |
| | *Mass Recovery* | | | | |
| Recovered amount of salt | 24% | 15% | 32% | 19% | 32% |
| Recovered amount of uranine | 25% | 29% | 33% | >14% | 34% |

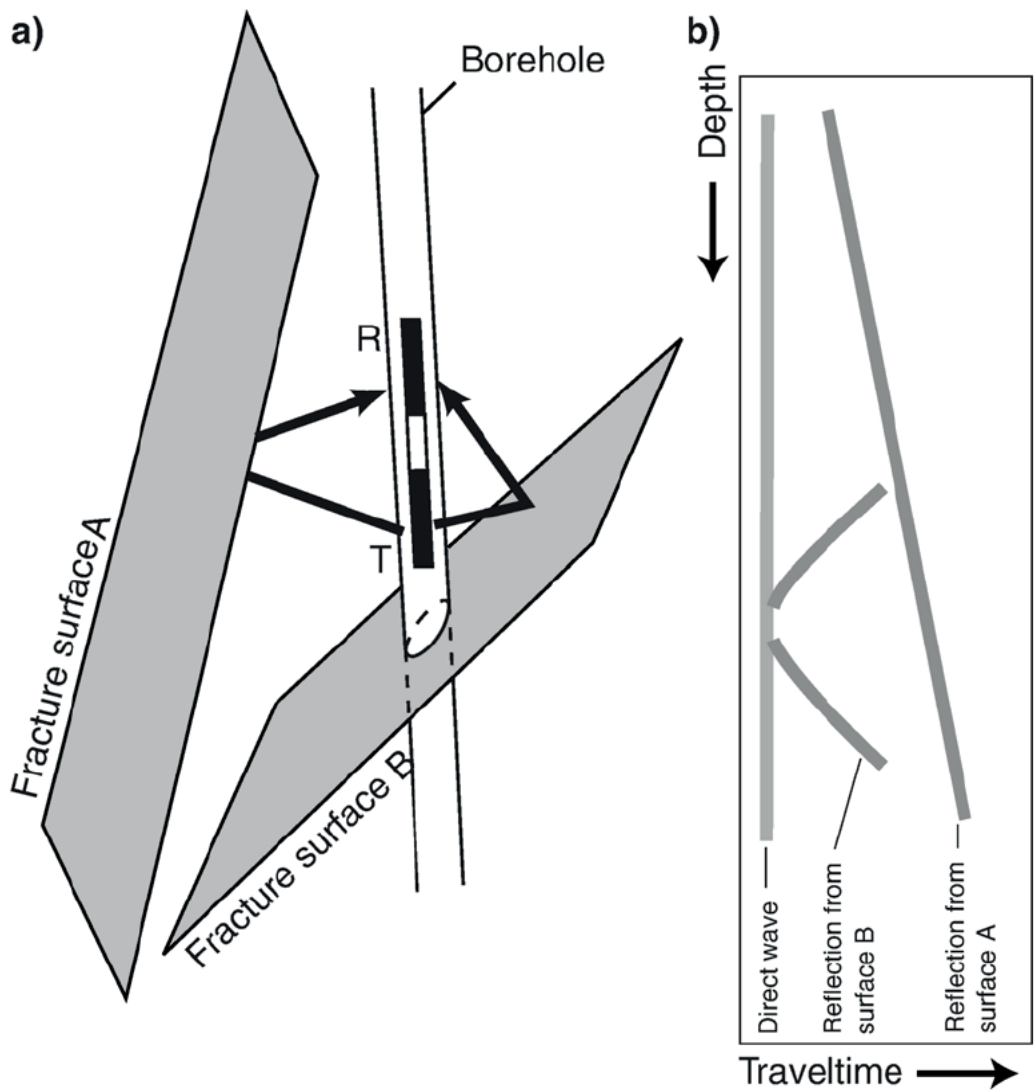

Figure 1. (a) Principle of the single-hole GPR reflection method, in which a transmitter T sends out a signal that is reflected and subsequently collected by a receiver R located in the same borehole as the transmitter. (b) Schematic reflection section illustrating typical reflection patterns arising from intersecting and non-intersecting fractures. Figure from Spillmann et al. [2007].

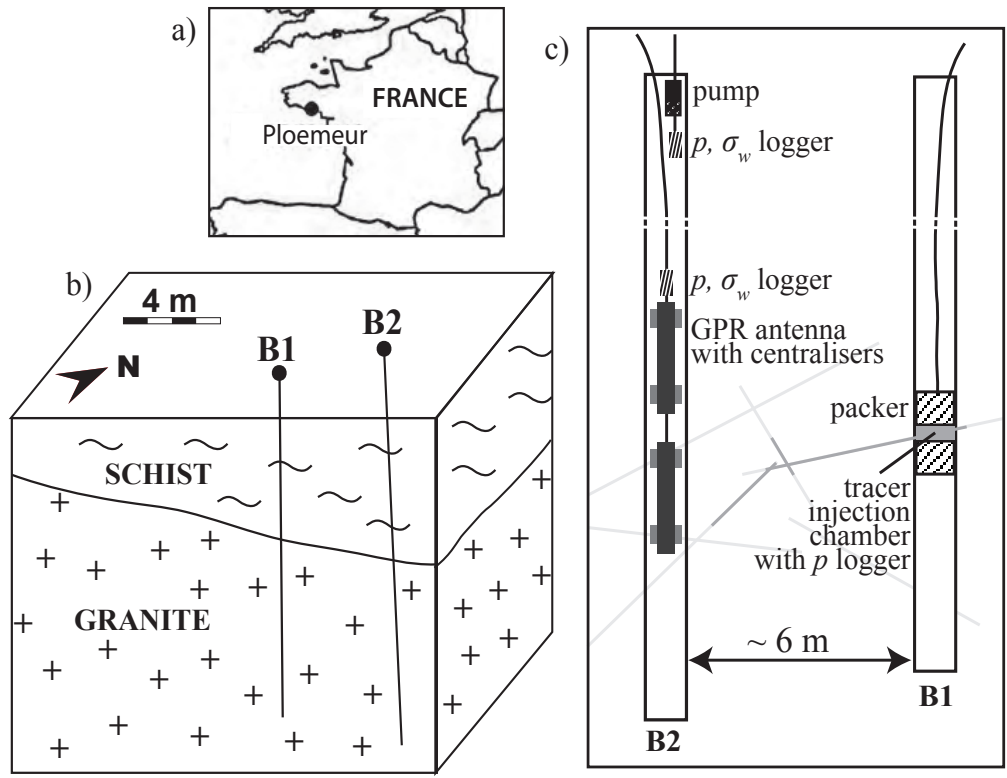

Figure 2. (a) Location of the Stang-er-Brune study-site in the vicinity of Ploemeur, France. (b) Geological model of the field site with a 30° dipping contact between micaschist and underlying granite. (c) Schematic of the data acquisition setup, in which p and σw logger refer to hydraulic pressure and groundwater conductivity loggers, respectively.

*Raw Data Section*     *Raw Reference Section*

Static corrections
Time zero correction
Geometry assignment
Geometrical scaling
Bandpass filter
Positioning correction
Wavelet filter
Wavelet normalization filter
Envelope scaling
Eigenvector filter
*Preprocessed Data and Reference Section*
*(Figures 4a and 4b)*

Difference
Division by smoothed envelope
*Relative Difference Section*
*(Figure 4c)*

(F-X deconvolution)
(Subtraction of running median)
*Filtered Difference Section*
*(Figure 4d)*

First-break mute
Kirchhoff depth migration
*Migrated Difference Section*
*(Figure 4e)*

Figure 3. Flowchart of the GPR processing steps with reference to figures showing intermediate results. Processing steps in parenthesis only apply to experiments Ia and Ib.

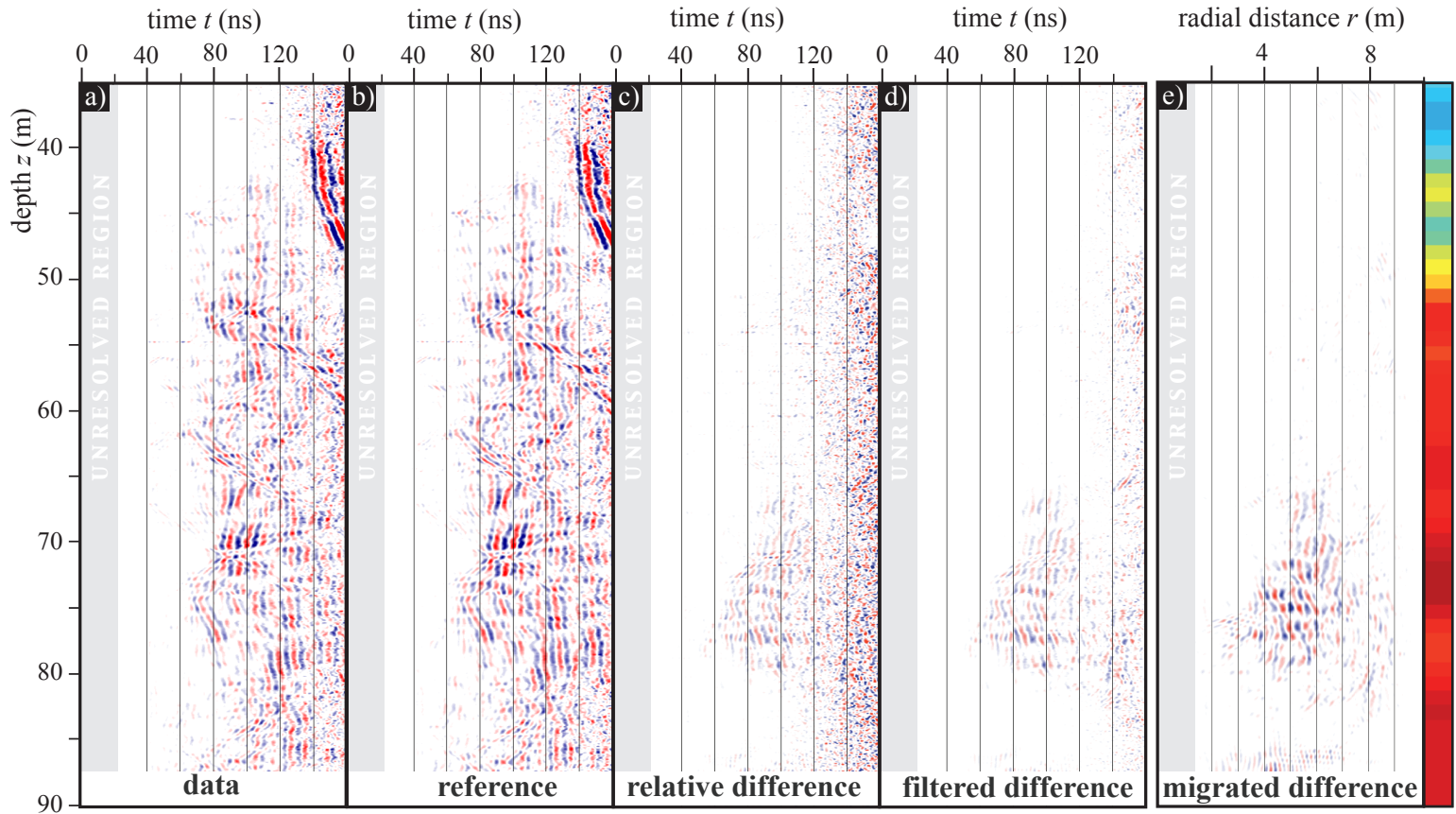

Figure 4. Results of data processing applied to a single-hole GPR data section of experiment Ib acquired after saline tracer injection at z = 78.7 m (tobs = 45 min, 4 m antenna offset). (a) Data and (b) reference section after preprocessing. (c) Relative difference between (a) and (b) normalized by the envelopes of (b). (d) As in (c), but after f-x deconvolution. (e) As in (d) but after depth migration using the velocity model shown to the right of the migrated difference. The axis aspect ratio r:z is 2:1.

$v$ (m/ns)

0.10    0.11    0.12

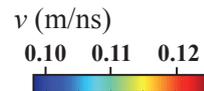

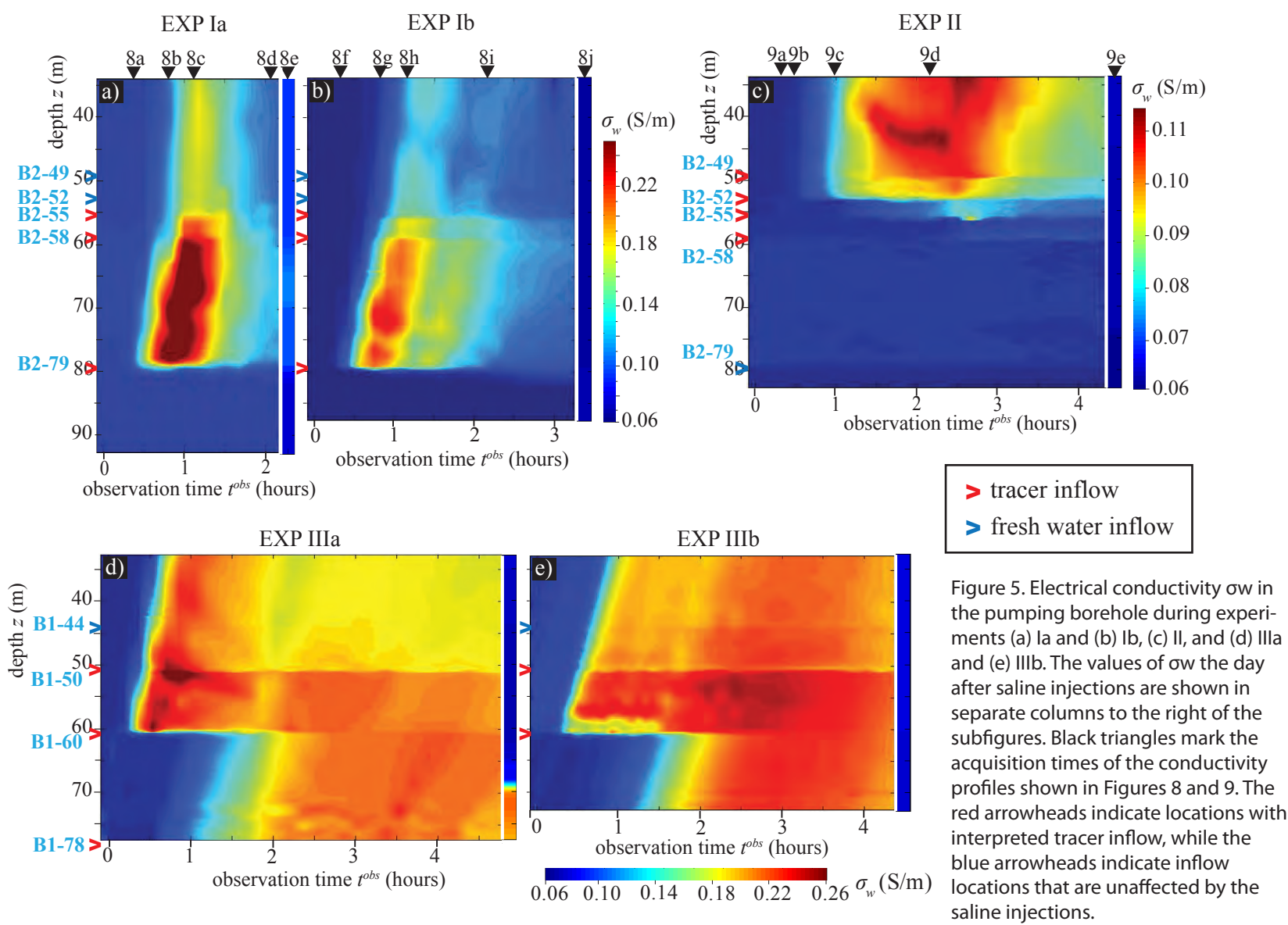

Figure 5. Electrical conductivity σw in the pumping borehole during experiments (a) Ia and (b) Ib, (c) II, and (d) IIIa and (e) IIIb. The values of σw the day after saline injections are shown in separate columns to the right of the subfigures. Black triangles mark the acquisition times of the conductivity profiles shown in Figures 8 and 9. The red arrowheads indicate locations with interpreted tracer inflow, while the blue arrowheads indicate inflow locations that are unaffected by the saline injections.

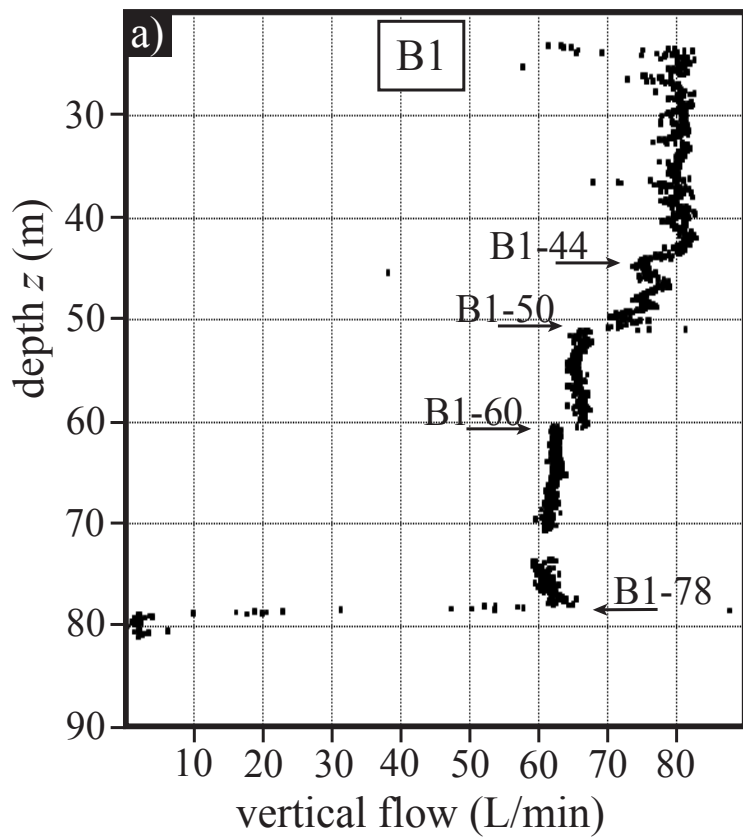

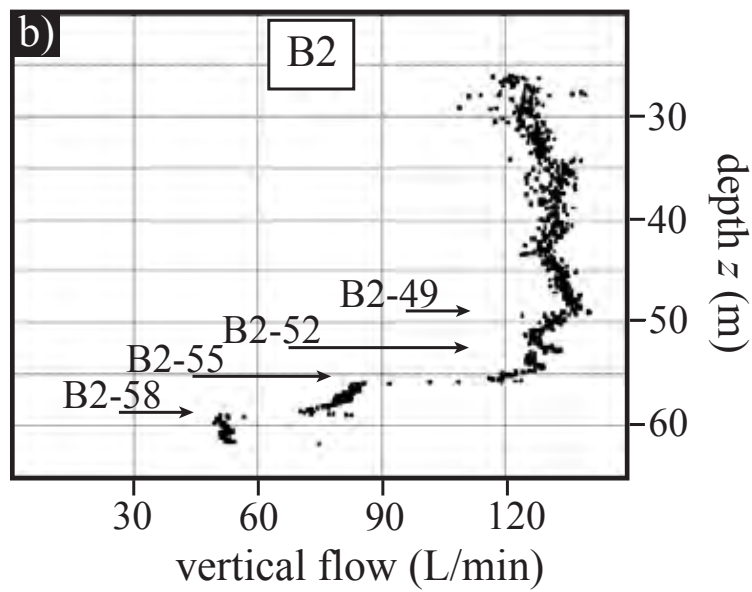

Figure 6. Induced vertical flow due to pumping in (a) B1 with 82 L/min and in (b) B2 with 138 L/min measured with an impeller flowmeter. All the flow below z = 60 m in B2 stems from fracture B2-79 at z = 79.3 m. Arrows indicate the locations of interpreted permeable fractures.

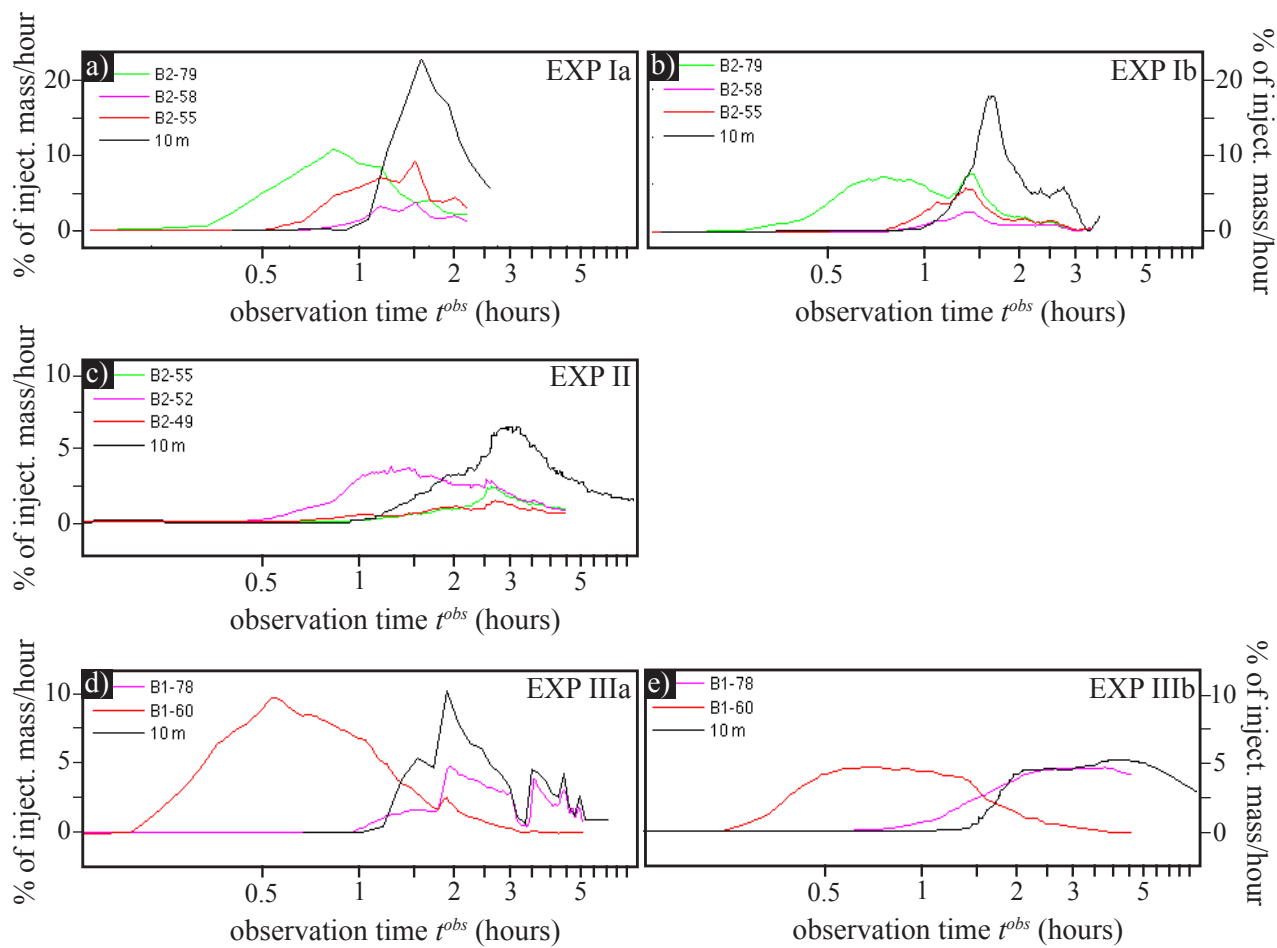

Figure 7. Local salt solute flux curves estimated at depth locations with significant tracer arrival (colored lines) and below the pump (z = 10 m, black line) normalized by the injected amount of tracer mass for experiments (a) Ia and (b) Ib, (c) II, (d) IIIa and (e) IIIb.

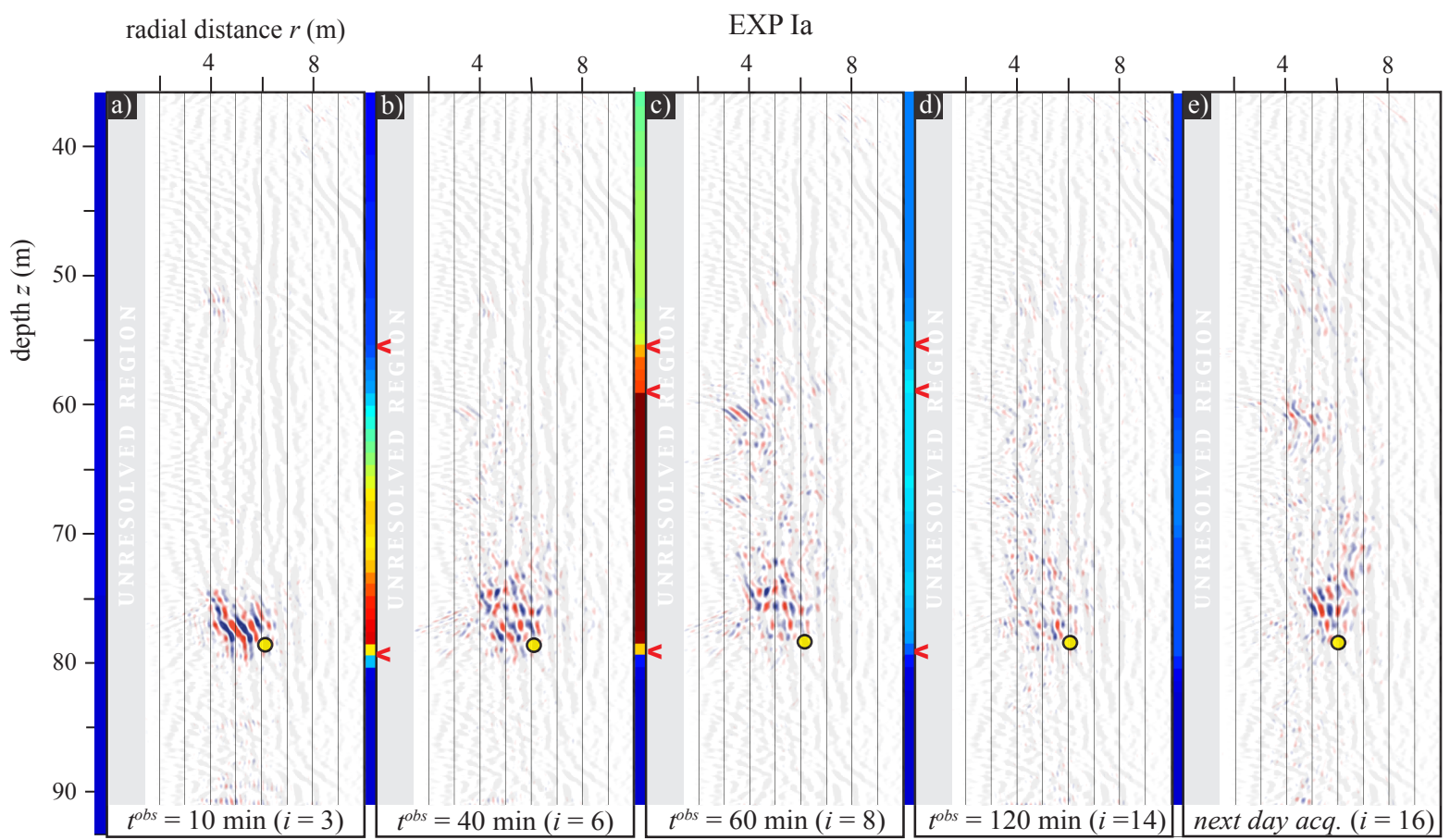

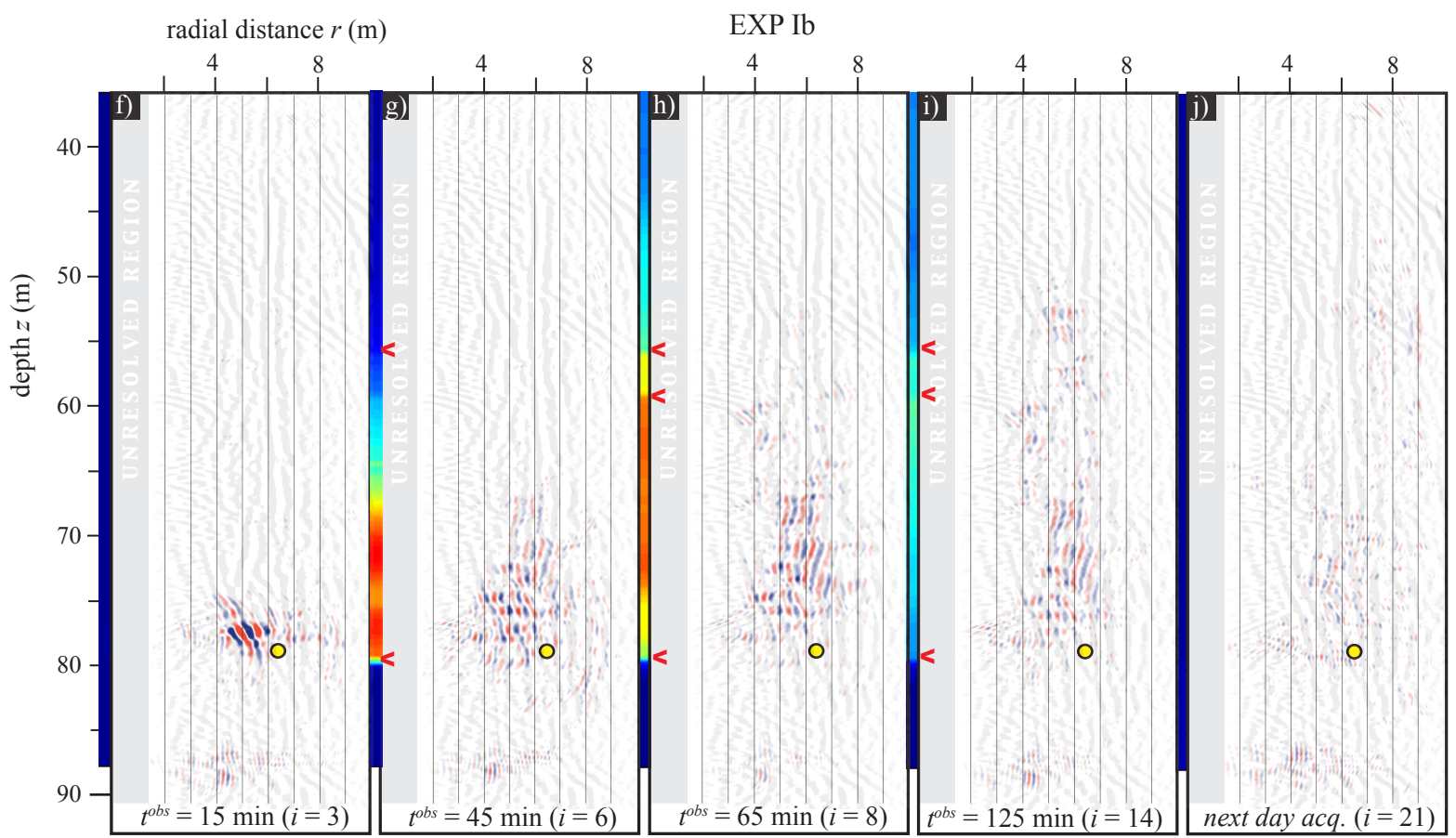

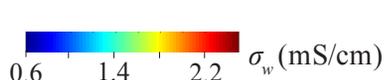

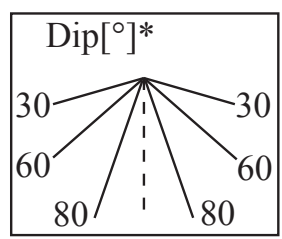

Figure 8. Migrated relative difference GPR sections acquired in B2 during experiment Ia (a-e) and experiment Ib (f-j) at different observation times tobs, superimposed on gray opaque GPR section that has been acquired under natural conditions [Dorn et al., 2012]. High difference patterns originate from increased salinity in fractures located at the front of each such pattern (i.e., the smallest radial distance r for each depth z). Note that we do not image any features at r < 1.5 m (gray region) because of the dominance of the direct wave at early times and its subsequent removal, which tends to remove superimposed reflections at early times. The corresponding electrical conductivities σw of the borehole fluid in B2 are shown in the color profile at r = 0 m.

Figure 9. Migrated relative difference GPR sections acquired in B2 during experiment II at different observation times tobs, superimposed on gray opaque GPR section that has been acquired under natural conditions [Dorn et al., 2012]. The corresponding electrical conductivities σw of the borehole fluid in B2 are shown in the color profile at r = 0 m.

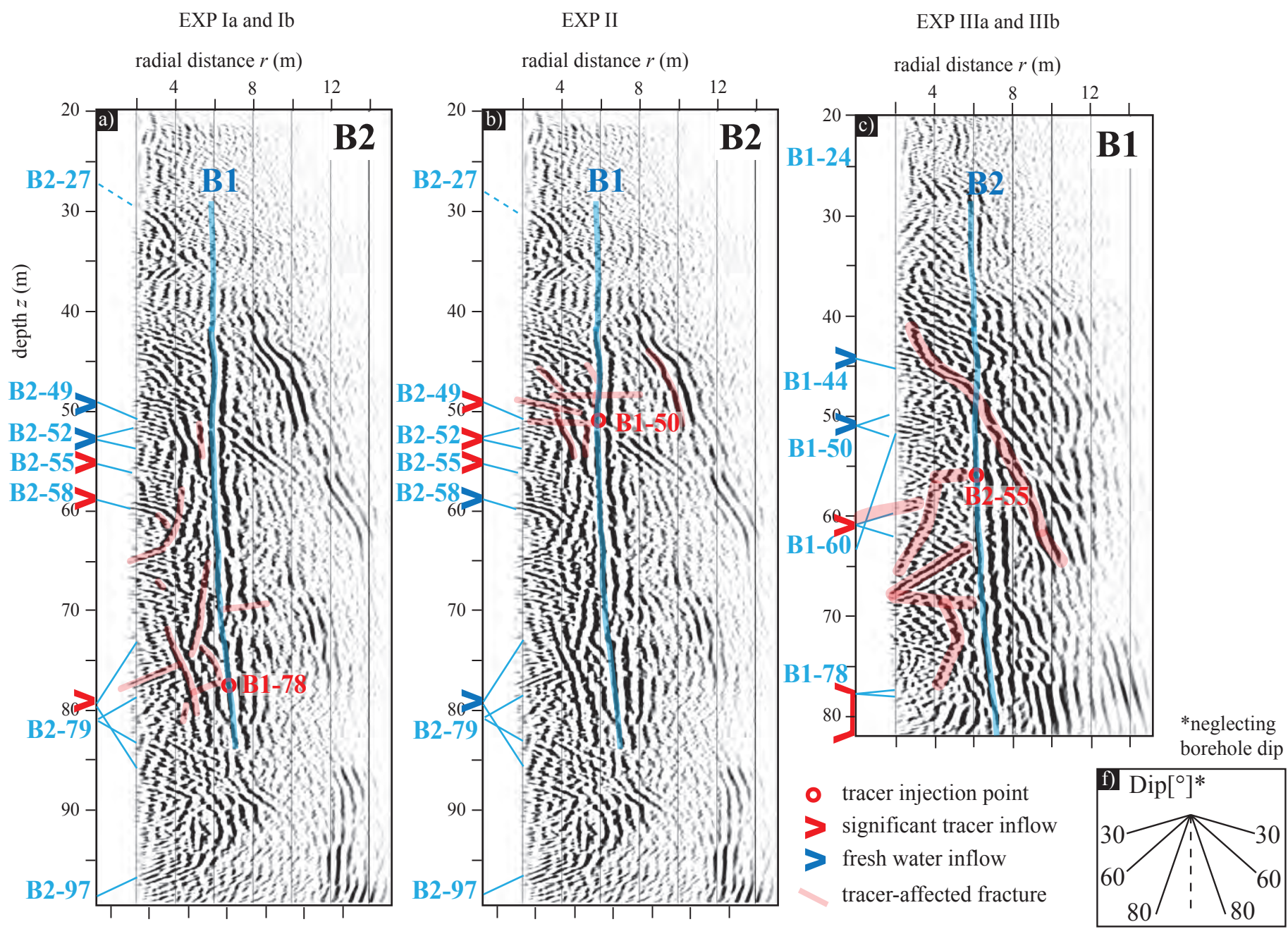

Figure 10. Extracts of the migrated multi-offset single-hole GPR sections of B1 and B2 from Dorn et al. [2012] with superimposed interpretations of tracer pathways for experiments (a) Ia and Ib, (b) II, (c) IIIa and IIIb. Red circles indicate the tracer injection points, while red and blue arrowheads locate saline and unaffected groundwater inflow into the pumping borehole, respectively. Light red regions highlight fractures through which the injected tracer is interpreted to move, whereas blue regions highlight reflections from other boreholes. Light blue letters refer to transmissive fractures identified in the boreholes using optical logs and flowmeter tests with corresponding blue lines indicating their corresponding dips [Le Borgne et al., 2007]. (f) Dip angles corresponding to the axis aspect ratio r:z of 2:1.

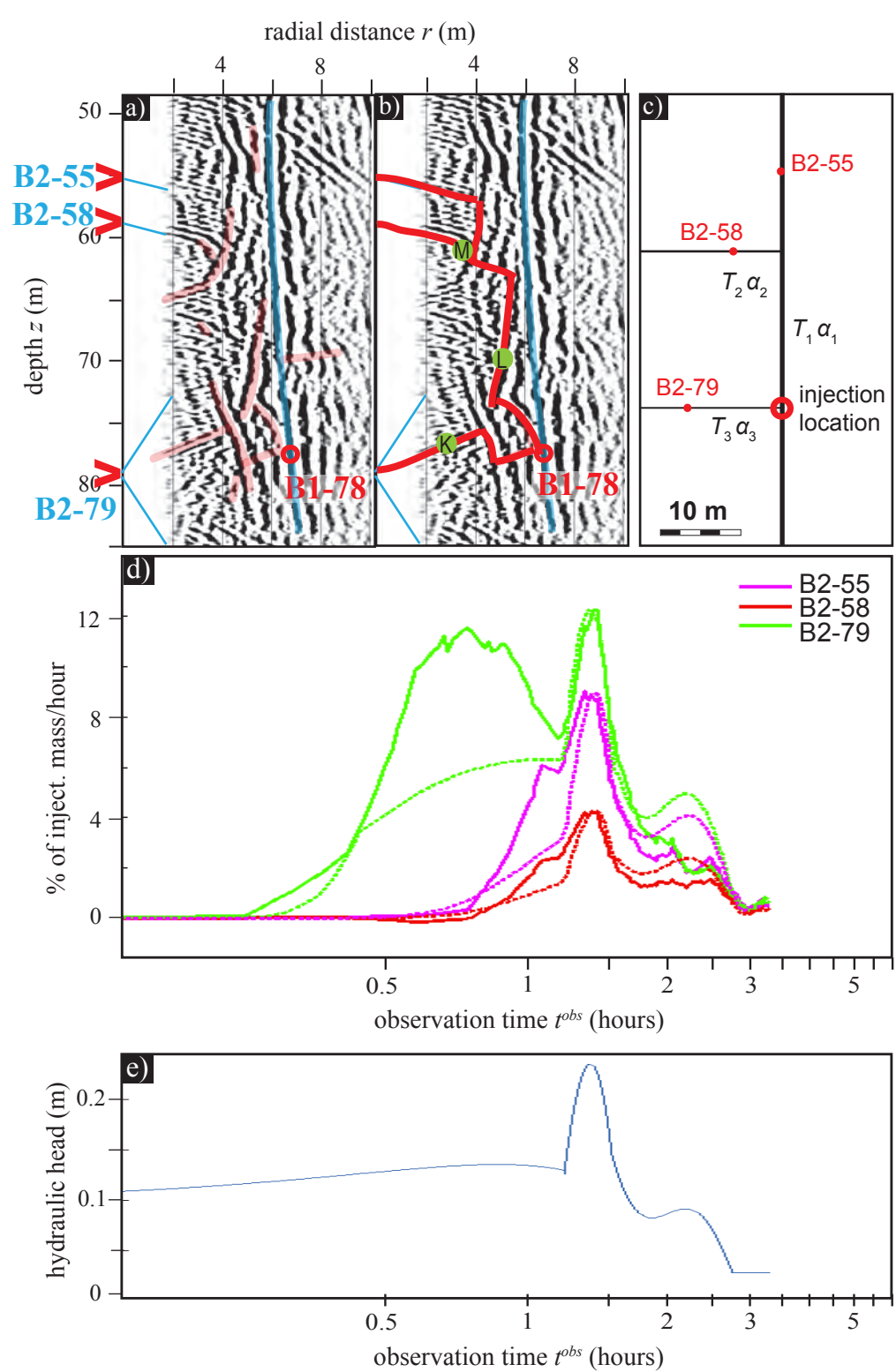

Figure 11. Transport model and modeling results for experiment Ib based on a simplified 3D fracture model with three intersecting rectangular fracture planes with an aperture of 1 mm. (a) Extract of Figure 10a, on which we assign (b) three continuous transport pathways between injection and outflow locations. Letters K-M in (b) refer to the positions considered in Figure 12. (c) Graph representing a simplified representation of the pathways in (b) used in the transport model to estimate effective hydraulic conductivities k and dispersivities α. (d) Observed (solid lines) and simulated (dashed lines) local salt flux curves at outflow locations B2-55, B2-58 and B2-79. The derived transport parameters are: k1 = 0.6 m/s, k2 = 2.3 m/s, k3 = 0.2 m/s, α1 = 0.6 m, α2 = 0.3 m and α3 = 0.2 m. (e) Hydraulic head used for the boundary conditions at the three outflow locations.

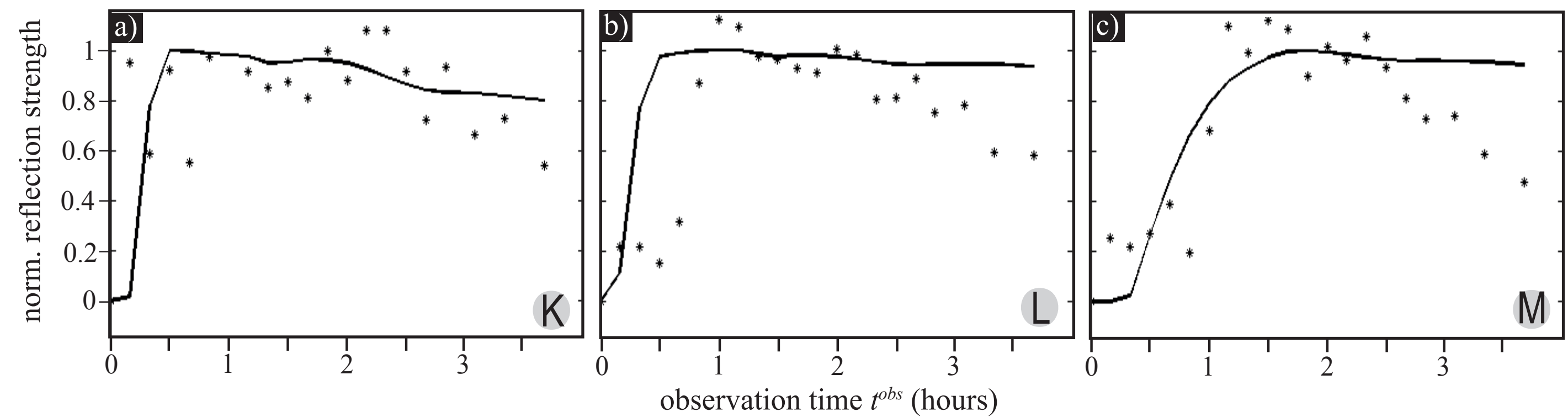

Figure 12. Time evolution of normalized reflectivity changes with respect to reference conditions observed in the GPR difference images (asterisks) and calculated from simulated tracer distributions (solid line) corresponding to locations (a) K, (b) L and (c) M in Figure 11b. At each location, we average the data over 1 m, but there is no smoothing over time.